\documentclass[letterpaper, 10 pt, conference]{ieeeconf}  

\IEEEoverridecommandlockouts                              
\usepackage{graphicx} 
\usepackage{amsmath} 
\usepackage{amssymb}  
\usepackage{multirow}

\title{\LARGE \bf
An Empirical Assessment of the Complexity and Realism of Synthetic Social Contact Networks*
}

\author{Kiran Karra$^{1}$, Samarth Swarup$^{2}$, and Justus Graham$^{1}$ 
\thanks{*Distribution Statement A: Approved for Public Release, Distribution Unlimited.  This work was supported in part by DARPA MAA Contract FA8750-17-C-0155 (NC082817-VT-GORDIAN); the views, opinions and/or findings expressed are those of the authors and should not be interpreted as representing the official views or policies of the Department of Defense or the U.S. Government; SS is additionally supported in part by DTRA CNIMS Contract HDTRA1-17-0118 and NIH Grant 1R01GM109718.}
\thanks{$^{1}$Hume Center,
        Virginia Tech, Arlington, VA 22203, USA
        {\tt\small www.hume.vt.edu}}%
\thanks{$^{2}$Biocomplexity Institute, Virginia Tech,
        Blacksburg, VA 20460, USA
        {\tt\small www.bi.vt.edu}}%
}

\begin{document}

\maketitle
\thispagestyle{empty}
\pagestyle{empty}

\begin{abstract}

We use multiple measures of graph complexity to evaluate the realism of synthetically-generated networks of human activity, in comparison with several stylized network models as well as a collection of empirical networks from the literature.  The synthetic networks are generated by integrating data about human populations from several sources, including the Census, transportation surveys, and geographical data. The resulting networks represent an approximation of daily or weekly human interaction. Our results indicate that the synthetically generated graphs according to our methodology are closer to the real world graphs, as measured across multiple structural measures, than a range of stylized graphs generated using common network models from the literature.

\end{abstract}

\section{Introduction}

Artificially generated graphs benefit from high demand in several application domains, wherever the phenomena of interest are driven by interactions between people, including health and medicine, communications, the economy, and national security. Lack of access to appropriate network data hampers the research community's ability to develop algorithms to analyze and gain insight from these transactional graph datasets.  Due to the access restrictions to real network data, there is value in crafting methods of synthetically generated data which faithfully represent behaviors of real world processes.   As such, many stylized methods for creating graphs with rigorously understood structural properties have been established, making collective steady progress towards better approximating structures of real world processes.  

Despite this progress, these relatively simple stylized methods aren't universally applicable and suffer from lack of realism for some applications.  We are particularly interested in creating realistic graphs which represent a complex set of interrelated processes involving a common subset of actors (i.e., the coherent alignment of disparate subgraphs which have many vertices in common and which represent different types of underlying activity).  We are also interested in the ability to embed various types of synthetic activities (subgraphs) into these larger ``background'' graphs to enable the advancement of subgraph detection algorithms on large and realistic datasets. This is also relevant to recent work on adversarial social network analysis~\cite{michalak17strategicSNA, waniek16hiding}, which could benefit from the application of realistic complex networks. Lastly, we are interested in graphs with high realism according to the particular domain of activity sought to be modeled by any given graph or collection of graphs. As a step towards achieving these objectives, we are examining the use of synthetic populations to generate realistic graphs. 

Synthetic population technology is widely used in a number of domains. It involves developing a highly detailed and data-driven representation of the entire population of a region. The synthetic agents are imbued with demographics, household structure, and activity patterns. The use of actual location data gives the synthetic populations a spatial structure that corresponds to the contingent spatial realities of the region being modeled. This leads to simulations that can be used to study specific phenomena as well as understand which aspects of a phenomenon generalize across regions. Synthetic social contact networks are constructed from synthetic populations by connecting synthetic agents who are spatiotemporally collocated in the model~\cite{EK+04}. From initial applications in the transportation domain~\cite{beckman97transims}, this approach has expanded to studies in economic analysis~\cite{axtell16firms, ballas06jobs}, epidemiology~\cite{parikh13transient, EG+04, swarup17heat}, land use planning~\cite{waddell02urbansim}, disaster response~\cite{parikh16jaamas}, and more~\cite{edwards09simobesity, wegman09alcohol}.

This has led to extensive evaluation and validation of the methodology and the synthetic population data that are generated~\cite{muller10population, guo07synthpop, elbers03poverty}. This is also detailed in Section~\ref{sec:synth_network_gen}. However there has been relatively little evaluation of the synthetic social contact networks that are generated from these populations (a few examples are~\cite{xia14comparison, EK+05, BE+04, xia13adequacy}).

The goal of this paper is to evaluate the synthetic population method for its ability to generate graphs that meet application needs of complexity and realism as described above.  This paper focuses on the ability of this method to represent the characteristics of real world data as compared to stylized graph generation methods. We evaluate synthetic social contact graphs, stylized graphs based on probabilistic graph generation, and real world data from online network data repositories according to a robust number of data-agnostic graph features.  Results are provided and suggestions for future research are outlined, particularly with regard to comparing against more nuanced datasets which represent complex behavior across multiple activity domains.

The rest of this paper is organized as follows.  Section II describes measures of graph complexity which were applied to all graphs to perform our analysis.  Section III describes our synthetic graph generation methodology and resulting instances of graphs.  Section IV describes the reference data used for comparison, including stylized graphs (classical methods) and real world graphs used for the analysis.  Section V describes the evaluation methodology and analytical results.  Finally, Section VI provides context for the results as a conclusion as well as several options for further research.

\section{Measures of graph complexity}\label{sec:measures_of_complexity}
In this section, we discuss the measures of graph complexity that we compute in order to compare graphs generated by different methods in a data agnostic fashion.  We compute eleven different measures of graph complexity, based on their efficacy to our problem, as well as their computability with respect to available open source packages.  The features we compute, and their respective definitions are stated below.

\begin{itemize}
	\item \textbf{Local Clustering Coefficient} $c_i = \frac{|\{e_{jk}\}|}{k_i(k_i-1)} : v_j, v_k \in N_i, e_{jk} \in E$, where $k_i$ is the out-degree of vertex-$i$, and $N_i = \{ v_j: e_{ij} \in E \}$ is the set of out-neighbors of vertex-$i$ \cite{clustering_coefficient}.  The local clustering coefficient of a vertex is a measure of the tendency of that node to cluster (i.e. be connected) to other nodes.
	\item \textbf{Page Rank} $PR(v) = \frac{1-d}{N} + d \sum_{u \in \Gamma^-(v)} \frac{PR(u)}{d^+(u)}$, where $\Gamma^-(v)$ are the in-neighbors of $v$, $d^+(w)$ is the out-degree of $w$, and $d$ is a dampling factor \cite{pagerank}.  Page rank is one measure of the ``importance'' of a node, where importance is proportional to the in-degree of each node.
	\item \textbf{Betweenness Centrality} $C_B(v) = \sum_{s \neq v \neq t \in V} \frac{\sigma_{st}(v)}{\sigma_{st}}$, where $\sigma_{st}$ is the number of shortest paths from $s$ to $t$, and $\sigma_{st}(v)$ is the number of shortest paths from $s$ to $t$ that pass through $v$.   \cite{betweenness}.  Betweenness is a measure of the centrality of the nodes in the graph; nodes that are considered to be more between have more shortest paths that pass through them than less between nodes.
	\item \textbf{Closeness Centrality} $c_i = \frac{1}{\sum_j d_{ij}}$, where $d_{ij}$ is the distance between nodes $i$ and $j$ \cite{closeness}.  Closeness is another measure of centrality, but differs from betweenness in that it is proportional to the sum of the shortest paths between the node of interest and all other nodes in the graph.
	\item \textbf{Vertex In Degree} is the number of incoming connections into each node of the graph.
	\item \textbf{Vertex Out Degree} is the number of nodes that this node connects to.
	\item \textbf{Von-Neumann Approximate Graph Entropy} - $H_{VN}(G) = \frac{|V|}{4} - \sum_{(u,v) \in E} \frac{1}{4 du dv}$, where $du = D(u,u) = \sum_{v \in V} A(U,v)$ \cite{bai2013graph}.  From Information Theory, entropy is a measure of randomness, and thus loosely quantifies the complexity of the graph from a probabilistic perspective.
	\item \textbf{Structure Connectivity} $C = E - V + 1$, where $E$ is the number of edges and $V$ is the number of vertices \cite{bonchev04yeast}.  The structure connectivity is an easily computable coarse measure of the density of the graph.
	\item \textbf{Structure Connectedness} $Conn[\%] = \frac{2E}{V(V-1)} \times 100$, where again $E$ is the number of edges and $V$ is the number of vertices \cite{bonchev04yeast}.  The structure connectedness is also an easily computable coarse measure of the density of the graph.
	\item \textbf{Average Intersite Distance} $\langle{d_i}\rangle = \frac{2W}{V}$, where $V$ is the number of vertices, and $W$ is the Wiener number of the network.  The Wiener number is defined as the sum of all distances in the network, and is computed by $W = \frac{1}{2}\sum_{i,j=1}^{V}d_{ij}$ \cite{bonchev04yeast}.  The average intersite distance is a measure of how easily one vertex can reach another vertex.
	\item \textbf{Vertex Distance Information} $I_r(vd) = \frac{1}{A \text{log}_2A} \sum_{i=1}^V a_i \text{log}_2 a_i$, where $a_i:\{ N_1(a_1), N_2(a_2), \dots, N_k(a_k) \}$ and $N_i(a_i)$ represents the number of vertices with vertex degree $a_i$ \cite{bonchev04yeast}.  The vertex distance information is a normalized measure of the entropy of the vertex degree distribution, and is another measure of the complexity of the graph.
\end{itemize}

For the clustering coefficient, page rank, betweenness, and closeness, we compute the defined feature for all nodes in the graph, and then sample the cumulative distribution function at probabilities $\left[ 0.10, 0.25, 0.50, 0.75, 0.90 \right]$ to capture the information given by the feature across all nodes in the graph. For the vertex degree computations, we compute the mean and standard deviation, and use both as features.  This yields a total of $34$ features per graph that are computed to compare the graphs with each other.

\subsection{Related work}

Earlier work on understanding the properties of synthetic social contact networks has taken two approaches. One is to develop mathematical approaches to understanding structural properties as they relate to properties of dynamics on the networks~\cite{EK+05, BE+04}. The other is to compare the synthetic social contact networks for different regions to understand the differences in their structural and dynamic properties~\cite{xia13adequacy, xia14comparison}.

Barrett et al.~\cite{BE+04} showed that there is a significant difference in how hard it is to shatter social networks as compared to infrastructural networks (transportation, power, and wireless radio networks), by examining the size of the largest component in the network as nodes are deleted in order of (residual) degree. Social networks were shown to be much more robust than the infrastructural networks.

Eubank et al.~\cite{EK+05} studied, among other things, the expansion characteristics of synthetic social contact networks. They showed that these networks have high expansion rates, which implies that contagious processes such as infectious diseases would spread very quickly on these networks. 

Xia et al. studied differences in social contact networks generated according to different methods~\cite{xia13adequacy} and the differences between the networks generated for two different cities (Delhi and Los Angeles)~\cite{xia14comparison}. They used both static and dynamic measures including degree distribution, temporal degree, vulnerability distribution, simulated epidemic curves, and the efficacy of two different epidemic mitigation interventions. Vulnerability is the probability that a node gets infected during an infectious disease outbreak. They showed significant differences due to the network generation methodology as well as across the two cities. 

Our contribution in this work differs from the earlier work in that we compute a larger set of structural features for the networks and do a comparison of synthetic social contact networks with real networks from the literature and also common stylized networks which are widely used in the literature.

\section{Synthetic network generation}\label{sec:synth_network_gen}


A synthetic social contact network is a representation of the interactions between people in a region due to collocation. To construct this network, we first generate a synthetic population for the region by combining data from multiple sources as described below. The main datasets used are listed in Table~\ref{table_synthpop}.

\begin{table}[h]
\caption{Data sets used for generating synthetic populations.}
\label{table_synthpop}
\begin{center}
\begin{tabular}{l|l}
\textbf{Data set} & \textbf{Description}\\
\hline
American Community Survey (ACS) & Demographics, \\
 & ~~~~household structure\\[2pt]
National Household Travel Survey (NHTS) & Daily and weekly \\
 & ~~~~activity patterns\\[2pt]
HERE (formerly Navteq) & Road networks and \\
 & ~~~~points of interest\\[2pt]
Dun \& Bradstreet (DnB) & Business locations\\[2pt]
National Center for & \\
~~~~Education Statistics (NCES) & School locations\\
\end{tabular}
\end{center}
\end{table}

\begin{figure*}[thpb]
  \centering
  \includegraphics[width=1\linewidth]{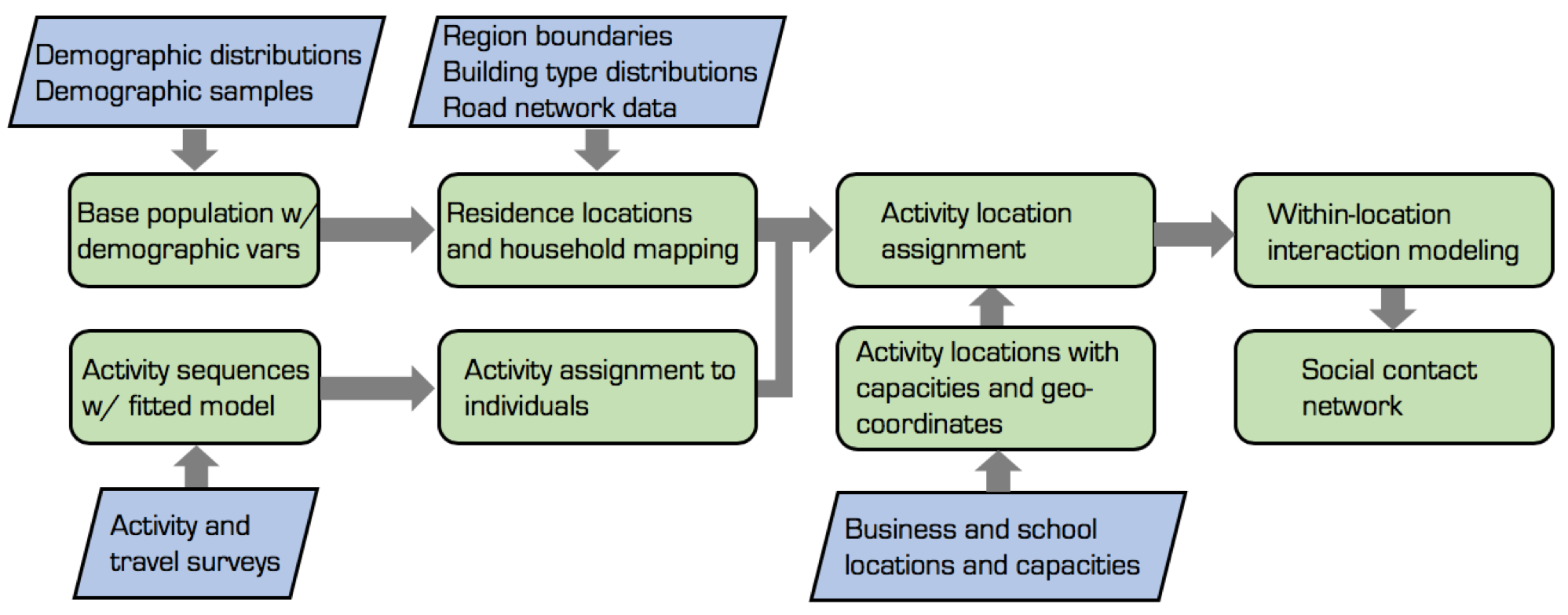}
  \caption{An illustration of the synthetic population and contact network generation methodology}
  \label{fig:synthpop_methodology}
\end{figure*}

The synthetic population generation process involves multiple steps. Detailed descriptions can be found elsewhere \cite{adiga15US}. We summarize each step below. The process is illustrated in Figure~\ref{fig:synthpop_methodology}. We generate the populations at the highest resolution allowed by publicly-available ACS data, which is the blockgroup. The Census Bureau does not release data at the block level publicly.

\subsection*{Base population generation}

The American Community Survey provides demographic data in two forms. One is in the form of distributions over key demographic variables in a blockgroup, such as householder age, household income, household size, and more. We henceforth refer to these as the marginal distributions. The other is an anonymized 5\% sample of the ACS records, known as the Public Use Microdata Sample (PUMS).

To generate a collection of synthetic individuals and households for the blockgroup whose demographics match the ACS data, we use a statistical procedure known as Iterative Proportional Fitting (IPF)~\cite{BBM96}. IPF creates a joint probability distribution over the chosen demographic variables by combining the marginal distributions with the PUMS sample for the blockgroup. This distribution is then sampled the needed number of times (equal to the number of households in the blockgroup) and the corresponding households are copied over from the PUMS sample to create the synthetic population.

\noindent
\textbf{Validation}: IPF provides a theoretical guarantee about the inferred joint distribution in a maximum entropy sense~\cite{ireland68contingency}. This means that the inferred joint distribution is as close as possible to the sample This step is further validated by comparing the distributions of variables in the synthetic population that were not included in the IPF procedure, with their corresponding marginal distributions as given by the ACS. 

\subsection*{Activity sequence assignment}

Once the base population has been generated, every synthetic individual is assigned an activity sequence for every day of a typical week. This is done using the National Household Travel Survey (NHTS) dataset. 

NHTS provides a set of travel records associated with demographics, from a nationally-representative sample of respondents. We group the activity types represented in the data into five categories: \emph{home}, \emph{work}, \emph{shopping}, \emph{other}, and \emph{school}. The last category also includes college activities.

NHTS travel records are matched with synthetic individuals using a two-stage fitted values approach~\cite{lum16fvm}. In the first stage, we calculate multiple summary statistics from the travel sequences, such as the fraction of the day spent in various activities (home, work, school, etc.). We then fit a regression to these summary statistics using the demographics as the independent variables.

In the second stage, this learned model is used to predict the same summary statistics for the synthetic households. We then find the closest NHTS survey household for each synthetic household and assign the corresponding activity sequences to the synthetic household. Closeness between synthetic and survey households is defined using, e.g., Mahalanobis distance between the fitted values of the summary statistics for the survey household and the predicted values for the synthetic households. The use of fitted values to evaluate the distance ensures that if two households have identical demographics, their distance is zero.

We repeat the method for each day of the week. At the end of this process, every synthetic individual has a meaningful weekly activity sequence assigned to it. Since the NHTS records include household structure, and we match an NHTS household to a synthetic household, intra-household correlations between activities are also preserved.

\noindent
\textbf{Validation}: Validation is done by comparing the aggregate distributions of time spent in various activities, broken down by demographics such as age and gender. For instance, the method reproduces the increasing trend of time spent at home with age (i.e., older people spend more time at home), as well as the gender difference between time spent at home (women spend more time at home on average). See~\cite{lum16fvm} for further details. 

\subsection*{Activity location assignment}

The previous step results in a synthetic population with demographics accurate to the blockgroup level and weekly activity sequences assigned to each individual. In this step, we assign geographical locations to each activity. This is divided into a procedure for home location assignment and a procedure for assigning locations for all other activity types.

To construct home locations for the synthetic households, we combine road network data from HERE with Census Bureau data on distributions of residential building types in each blockgroup and the blockgroup shapefiles.

Residence locations are constructed along the sides of the roads within the blockgroup. The kind of residence (single-family home, duplex, small or large apartment building) is chosen based on the road category. For instance, single-family residence locations are chosen along smaller roads and larger apartment buildings are chosen alongside larger roads. Synthetic households are allocated residences iteratively until all households have been assigned a home location.

Once home locations have been assigned, locations for other activity types are assigned using a gravity model~\cite{erlander90gravityModel}. Locations are first assigned for the so-called anchor activities (work and school). The gravity model chooses locations with probabilities proportional to their capacities and inversely proportional to the square of the distance. To make the computation tractable, a cutoff of 50 miles is assumed.

Then the locations for shopping and other activities are chosen, taking into account the locations of the anchor activities both before and after these activities. Thus, if a synthetic person's activity schedule involves going shopping between work and home activities, the shopping location is likely to be chosen inbetween the work and home locations.

\medskip
\noindent
\textbf{Validation}: This step has been validated in multiple ways. The most direct validation has been through estimating trip counts between traffic analysis zones and comparing with data~\cite{EG+04}. Beckman et al. also validated the mobility model indirectly by developing a model of cell phone network traffic and comparing measure of the traffic with data~\cite{beckman13spectrum}. They also showed that the distribution of distances traveled is a power law, which matches other estimates in the literature~\cite{gonzalez08mobility}.

\subsection*{Social contact network construction}

The synthetic population generation methodology results in a bipartite person-location network where the edges indicate the locations that the persons visit, annotated with the start-time and duration of the visit. In what follows below, we are using a synthetic social contact network for Montgomery County, VA, USA, which is openly available~\cite{swarup14challenge}.

\begin{figure}[thpb]
  \centering
  \includegraphics[width=1\linewidth]{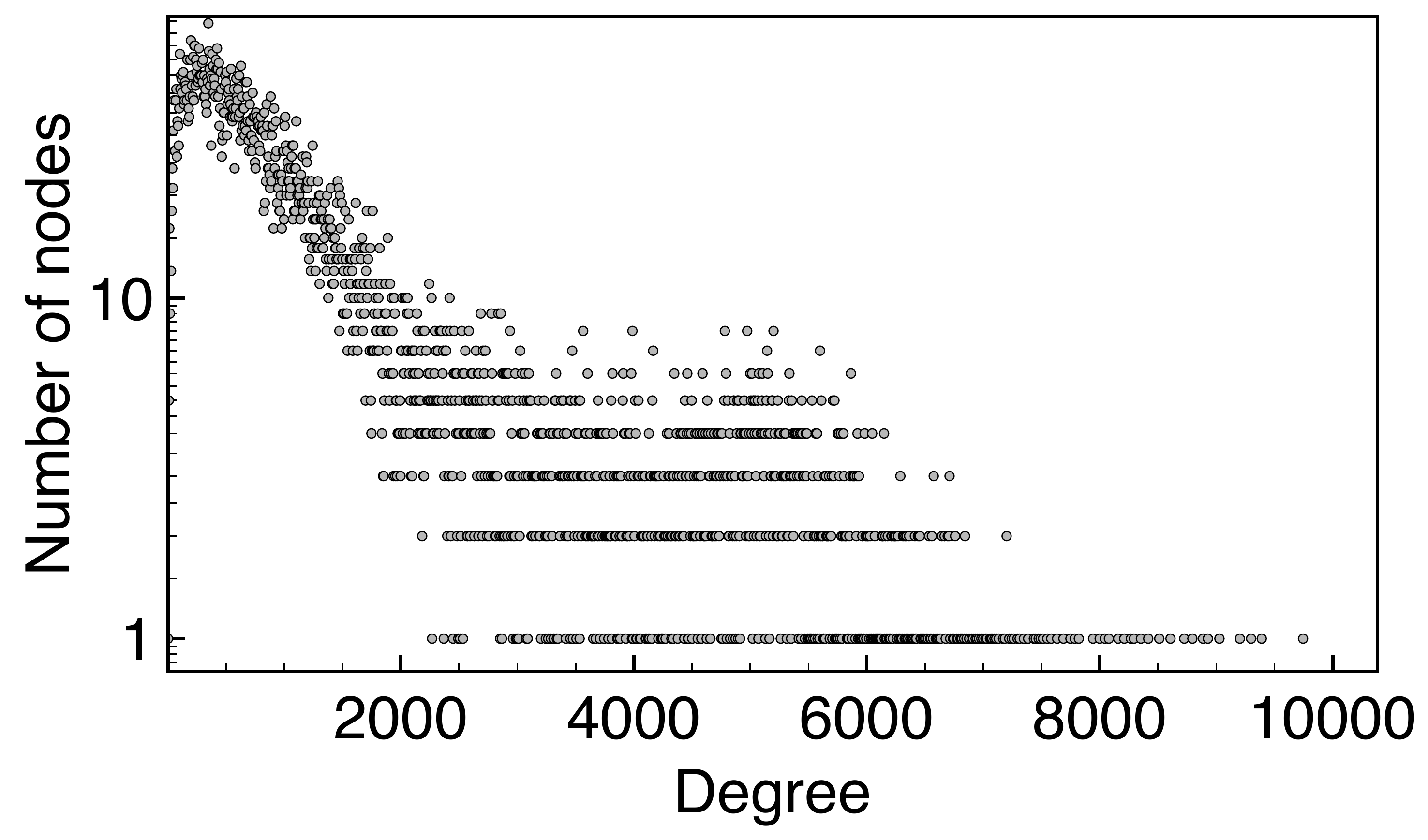}
  \caption{Degree distribution of the social contact network, assuming that all persons who could come into contact with each other based on collocation, actually do.}
  \label{fig:degree-distribution}
\end{figure}

From the bipartite network, we extract the induced person-person social contact network, where two persons might have an edge between them if they are at the same location for an overlapping period of time. The resulting network containing 63129 nodes and 45544774 edges, giving us an average degree of 1442.92. The degree distribution is shown in Figure~\ref{fig:degree-distribution}.

\begin{figure}[thpb]
  \centering
  \includegraphics[width=1\linewidth]{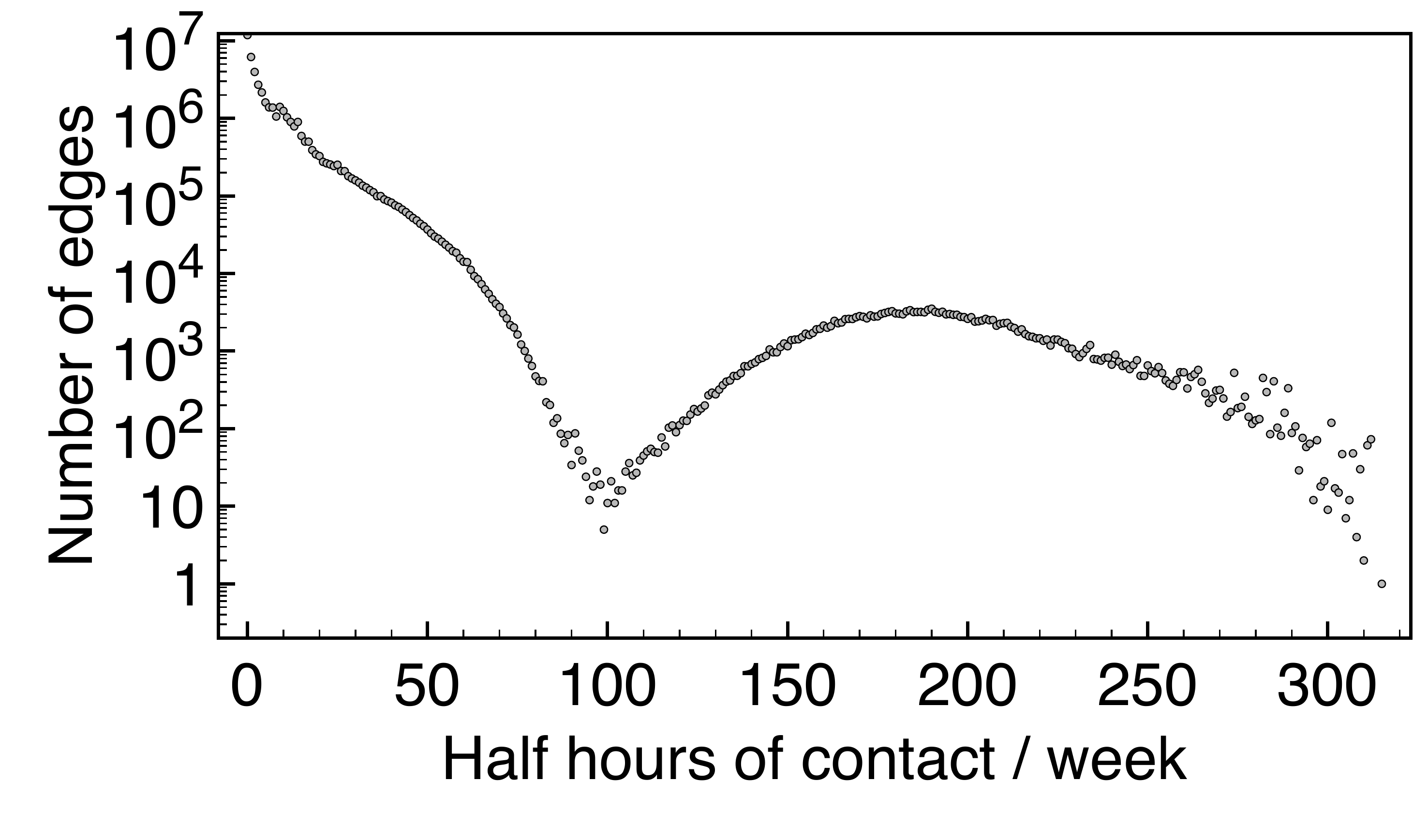}
  \caption{Distribution of contact durations, rounded to the nearest half hour.}
  \label{fig:duration-distribution}
\end{figure}

We also plot the distribution of durations of contact in this network, rounded up to half-hour intervals. This is shown in Figure~\ref{fig:duration-distribution}. This shows us that there are two distinct regimes of interaction in this model, with an inflection point at 50 hours of contact per week.

However, it is reasonable to assume that not all persons who visit a given location come into contact with each other. For example, we tend to interact with a subset of co-workers within the same building. If we go to a shopping mall, we actually interact with very few other people who are there. Therefore, in what follows, we assume that the probability that two persons come into contact depends on the kind of activity they are engaged in. We introduce five probabilities, $P\_HOME$, $P\_WORK$, $P\_SHOPPING$, $P\_OTHER$, $P\_SCHOOL$, corresponding to the probability that an induced edge in the person-person network corresponding to that activity is kept. We use different values for these probabilities to generate multiple social contact networks and examine their structure.

If all probabilities are 1.0, then all the induced edges are kept, and we obtain the network above. However, if we assume probabilities to be, e.g., $\{1.0, 0.01, 0.01, 0.01, 0.1\}$ (in the order the variables are listed above), we obtain a network with 2559061 edges, with an average degree of 81.17.

The degree distribution and contact duration distribution for this network are shown in Figures~\ref{fig:degree-distribution-prob4} and \ref{fig:duration-distribution-prob4} below.

\begin{figure}[thpb]
  \centering
  \includegraphics[width=1\linewidth]{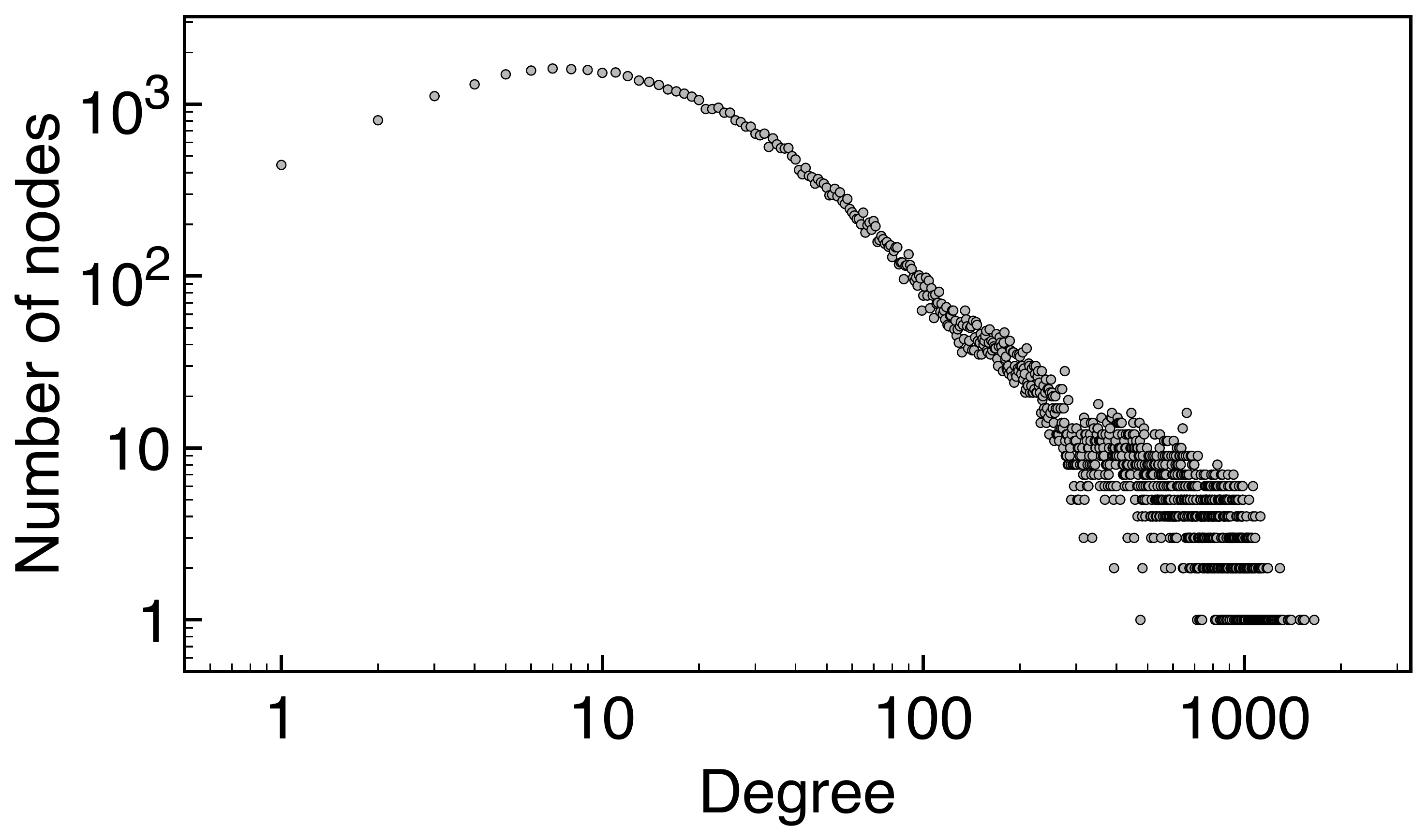}
  \caption{Degree distribution of the social contact network, assuming contact probabilities to be \{1, 0.01, 0.01, 0.01, 0.1\} for home, work, shopping, other, and school activities, respectively.}
  \label{fig:degree-distribution-prob4}
\end{figure}

We see that while the contact duration distribution looks qualitatively similar to the full network, the degree distribution now exhibits a power-law-like tail (Figure~\ref{fig:degree-distribution-prob4} is plotted on a log-log scale), while the degree distribution for the full network has an exponential character (Figure~\ref{fig:degree-distribution} is plotted on a log-linear scale).

\begin{figure}[thpb]
  \centering
  \includegraphics[width=1\linewidth]{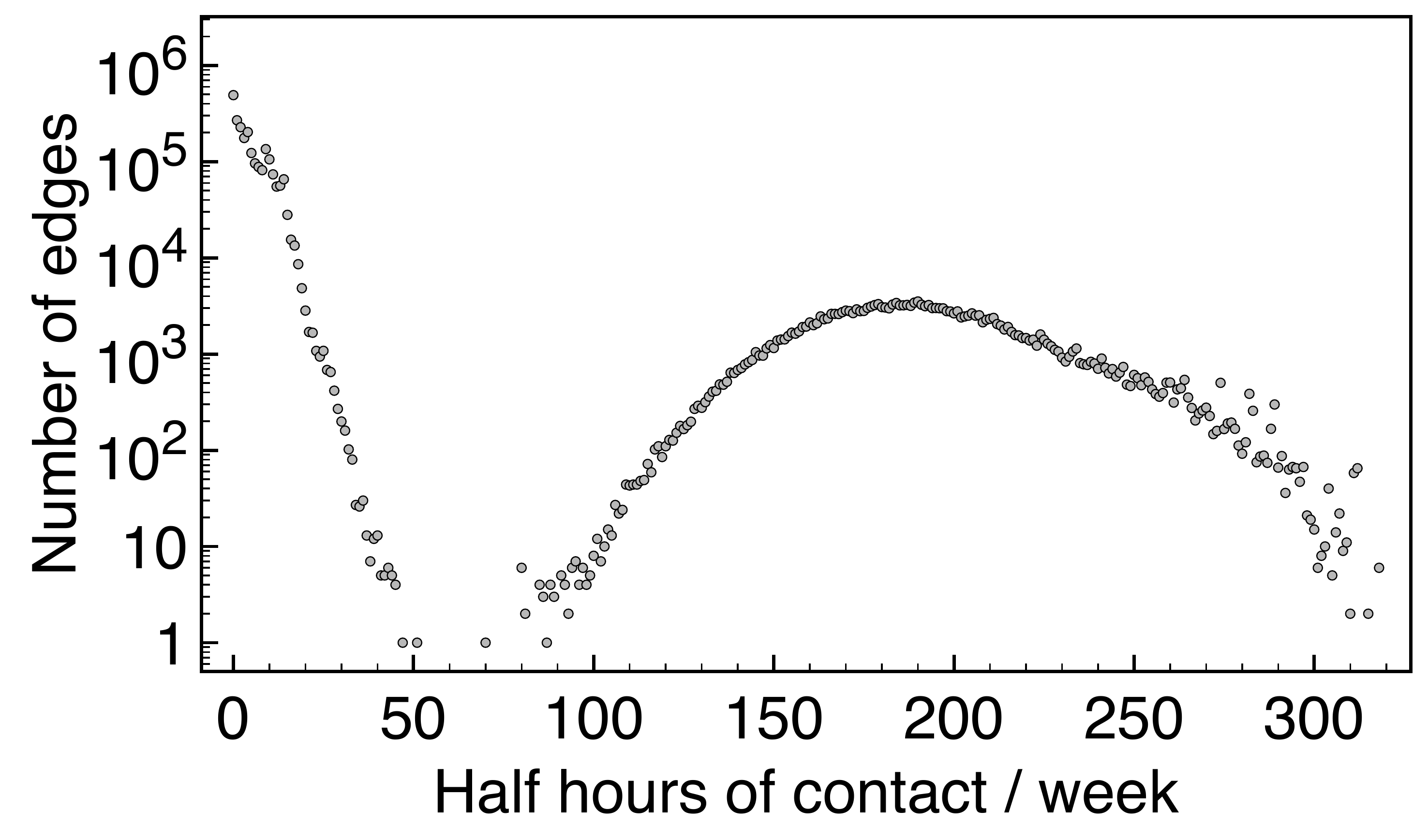}
  \caption{Distribution of contact durations, rounded to the nearest half hour for the social contact network with contact probabilities \{1, 0.01, 0.01, 0.01, 0.1\} for home, work, shopping, other, and school activities, respectively.}
  \label{fig:duration-distribution-prob4}
\end{figure}

Thus, by sampling edges representing different activity types with different probabilities, we can generate multiple social contact networks that exhibit different structural properties. Table~\ref{table_synth_networks} lists the networks we have generated using different sampling probabilities.

In addition to the full induced person-person network, we create networks which have each activity type in isolation. We also create networks with reasonable choices of the probabilities, denoted as $G_1 - G_4$ in Table~\ref{table_synth_networks}.

\begin{table}[h]
\caption{Synthetic social contact networks. Probabilities are listed in the order home, work, shopping, other, school.}
\label{table_synth_networks}
\begin{center}
\begin{tabular}{c|c|c}
\textbf{Name} & \textbf{Probabilities} & \textbf{Avg. degree}\\
\hline
Full & \{1, 1, 1, 1, 1\} & 1442.92\\
Home & \{1, 0, 0, 0, 0\} & 3.7 \\ 
Work & \{0, 0.01, 0, 0, 0\} & 17.83 \\ 
Shopping & \{0, 0, 0.01, 0, 0\} & 7.36 \\ 
Other & \{0, 0, 0, 0.01, 0\} & 6.28 \\ 
School & \{0, 0, 0, 0, 0.01\} & 17.29 \\ 
$G_1$ & \{1, 0.01, 0, 0, 0.01\} & 26.11 \\ 
$G_2$ & \{1, 0.01, 0.01, 0, 0.01\} & 31.02 \\ 
$G_3$ & \{1, 0.01, 0.01, 0.01, 0.01\} & 35.69 \\ 
$G_4$ & \{1, 0.01, 0.01, 0.01, 0.1\} & 81.17 \\ 
\end{tabular}
\end{center}
\end{table}

\section{Graphs for Comparison}

Three broad groups of network generation methods are compared: synthetic social contact networks described in Section~\ref{sec:synth_network_gen}, stylized graphs, and real world networks.

\subsection{Stylized Networks}
A collection of stylized networks were chosen to exercise structural variety among the various metrics chosen for evaluation.  These network types are trivially generated in extant graph analysis software packages such as NetworkX~\cite{team2014networkx}.  In order to provide a meaningful basis for comparison, these graphs were generated with parameters chosen according to Table~\ref{stylized_networks_parameters} to ensure that average degree fell within a similar range to the synthetic social contact networks as well as the real world data chosen from various domains.  Network size (i.e. node count), $n$, was chosen to provide computational tractability while also providing sufficient depth to accommodate a range of average degrees found in comparable real world networks.  Of note, while the node count in stylized networks is greater than in real world networks, these sizes were chosen to provide enough data for robust statistics; node count also would not affect the metrics under consideration, ceteris paribus.  In total, there are three average degree settings used for the four different stylized graph generation methods, yielding twelve total stylized graphs.  

\begin{table}[h]
\caption{Stylized Networks Generated Parameters.}
\label{stylized_networks_parameters}
\begin{center}
\begin{tabular}{c|c|c|c|c}
\textbf{Avg. Degree} & {} & \textbf{15} & \textbf{40} & \textbf{65} \\
\hline
\multirow{2}{*}{\textbf{\textit{Erdos-R\'enyi}}}&n&10000&10000&10000\\&p&0.0015&0.0004&0.0065\\
\hline
\multirow{3}{*}{\textbf{\textit{Newman-Watts}}}&n&10000&10000&10000\\&k&10&27&43\\&p&0.50&0.54&0.50\\
\hline
\multirow{2}{*}{\textbf{\textit{Random-Regular}}}&n&10000&10000&10000\\&d&7&20&32\\
\hline
\multirow{3}{*}{\textbf{\textit{Powerlaw-Cluster}}}&n&10000&10000&10000\\&m&8&20&33\\&p&0.50&0.50&0.50\\
\end{tabular}
\end{center}
\end{table}

\textbf{\textit{Erdos-R\'enyi}} graphs are widely used for the study of random graphs in the field of graph theory; they are constructed simply by defining the number of nodes $n$ and assigning a constant probability $p$ for all node pairs, which defines the probability of an edge occurring between two randomly selected nodes. While foundational, Erdos-R\'enyi graphs tend not to structurally achieve the degree distribution often found in real world graphs.  More specifically, the degree distribution follows a Poisson, while real world graphs tend to follow a power law~\cite{ornl_survey}; thus other methods are included in this analysis.

\textbf{\textit{Newman-Watts}} graphs modify the nature of random chance by mimicking ``small world'' behaviors found in real social networks, thereby emphasizing local structure.  Newman-Watts graphs raise the probability of two nodes being connected in certain locales while suppressing the probability of two nodes chosen at random being connected \cite{newmanwatts}.  The parameters defining the Newman-Watts model are $n$, the number of nodes in the network, $k$, the number of neighbors that each node is connected to in a ring topology, and $p$, the probability of adding a new random (or shortcut) edge for each node.

\textbf{\textit{Random-Regular}} graphs are bound to the restriction of being regular, such that each node has the same exact degree; this differs from the Erdos-Renyi model in which the distribution follows a Poisson model \cite{ornl_survey}.  The parameters defining the Random-Regular model are $n$, the number of nodes in the network, and $d$, the degree of each node, under the constraint that $n \times d$ be an even number.

\textbf{\textit{Powerlaw-Cluster}} graphs are constructed to preserve the notion found in real world networks regarding the existence of ``hubs'', which while rare have enormous degree.  The specific implementation of this graph combines the aforementioned power-law degree distribution (the existence of high degree hubs) with the local clustering notions from the Watts model above \cite{holme2002growing}.  The parameters defining the Powerlaw-Cluster graphs are $n$, the number of nodes in the network, $m$, the number of random edges to add for each new node, and $p$, the probability of adding an edge to form a triangle after adding an initial random edge.

\subsection{Real World Networks}
Real world networks fall along two broad categories: social and other.  Given that the synthetic social contact graphs detailed in Section~\ref{sec:synth_network_gen} are most closely related to social interaction, it is reasonable to overweight social networks in selecting real data.  Notably, the average degree of these social networks spans a wide range, from 7 to 209, yet these social graphs (and others available in the open) tend to the range from 15 to 75, thus the basis for parameterizing stylized graphs at the intervals in Table~\ref{stylized_networks_parameters}.  The analysis includes other graphs as well for diversity of analysis based on type of underlying activity, as well as much larger average node degree, as can be seen in the economic datasets.  A brief description of the real world networks used in this evaluation are shown in Table~\ref{real_world_networks}.  

\begin{table}[h]
\caption{Real World Network Parameters. \protect\\$\overline{\mathbf{D}}$ is the average degree. }
\label{real_world_networks}
\begin{center}
\begin{tabular}{c|c|c|c|c}
\textbf{Name} & \textbf{Type} & \textbf{Vertices} & \textbf{Edges} & $\overline{\mathbf{D}}$ \\
\hline
econ-beaflw & other & 507 & 53K & 209 \\
econ-beause & other & 507 & 44K & 174\\
piston & other & 2K & 98K & 96\\
email-EU-core & social & 1005 & 25571 & 50\\
p2p-Gnutella08 & social & 6301 & 20777 & 7\\
\end{tabular}
\end{center}
\end{table}

\section{Evaluation Methodology, Results, and Discussion}

\begin{figure}[thpb]
  \centering
  \includegraphics[width=1\linewidth]{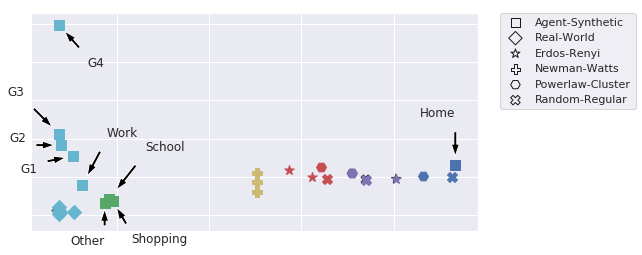}
  \caption{We show the 34 dimensional representation of the graph reduced to two dimensions using PCA, where the marker types represent the true underlying graph generation mechanism as shown in the associated legend, and the colors represent the K-Means clusterer's prediction of the type of graph generation mechanism.  It is seen that the clusterer cannot distinguish between the multichannel Agent-Synthetic graphs, G1 - G4, from real world network data.  All stylized graphs are easily distinguished from the Agent-Synthetic approach, although they are confused amongst themselves.}
  \label{fig:cluster_results}
\end{figure}

To compare the five empirical graphs, twelve stylized graphs, and eight synthetic social contact graphs, we chose a heuristic set of features which sought to provide robust characterizations as well as computationally tractability.  The features that were computed on each graph were described above in Section~\ref{sec:measures_of_complexity}, using the \textbf{graph-tool} computational framework~\cite{peixoto_graph-tool_2014}.  After computing the features, clustering was performed in order to examine whether the generated graphs could be separated based on the computed features of each graph.  K-Means clustering was used, with the number of clusters equaling the number of graph types that were generated, in this case, six~\cite{kmeans}.  The features were then reduced to two dimensions using PCA as the dimensionality reduction~\cite{pca} for visualization purposes.  Figure~\ref{fig:cluster_results} represents the results of the clustering and PCA.  In it, the various colors represent the K-Means clustering output, while the various marker shapes represent the true graph type, from a graph generation perspective.  

The key results indicate that the synthetic population-based graphs (referred to as Agent-Synthetic in Figure~\ref{fig:cluster_results}) are generally more similar to the real-world graphs than the stylized graphs.  Within each of the three broad graph classes, measures of similarity vary: the synthetic graphs (of both classes) tend to be more broadly distributed while the real world graphs are narrowly distributed.  In fact, the real-world graphs, despite having an enormous variance of average degree, are clustered very tightly in Figure~\ref{fig:cluster_results}.  Also of note, while the stylized graphs are pushed away from the real world graphs almost entirely in the X dimension, the agent-synthetic graphs gain distance almost entirely in the Y dimension.  We suggest this Y-axis distribution needs to be accounted for through additional interpretation, but we can conclude that the agent-synthetic graphs provide a more advantageous approximation to the real world graphs than the stylized counterparts.

Additional nuance should be used when evaluating these results, based on the underlying type of activity that is being modeled.  Within the agent-synthetic networks, it is seen that the graphs labelled Work, School, Other, Shopping, and Home are actually models of only a single type of underlying transaction.  The real world networks also only model a single type of underlying transaction.  G1-G4 graphs however jointly model the same entities performing different types of transactions all within the same graph.  Moreover, as one can see from Table~\ref{table_synth_networks} that G1 models the fewest channels as the third and fourth elements (representing Shopping and Other, respectively) are zeroed out.  As we transition from G1 to G4 in ascending order, either more elements become non-zero, or, the value of an element is raised.  For example, from $G1 \rightarrow G2$, the third element goes from $0$ to $0.01$.  From $G2 \rightarrow G3$, the fourth element goes from $0$ to $0.01$.  From $G3 \rightarrow G4$, the fifth element goes from $0.01$ to $0.1$.  These increases correspond to magnitude along the Y axis in this high dimensional space.

Summarily, when the underlying graph is normalized to represent only single underlying activity types, then the agent-synthetic data provides a better representation of real data from similar domains (e.g. social network data).  As the agent-synthetic data accumulate greater complexity through adding additional underlying activity types (i.e. simulate a combination of resources into a coherent graph), the distance grows along the Y axis, implying that while the data is less similar to the real world networks, it is dissimilar in a way different than the stylized networks.  

A final observation is that while the distance between the real world graphs and G1 - G4 is greater than with the individual networks Work, School, etc., the clustering algorithm assigned G1 - G4 into the same cluster as the real world graphs, as shown in Figure~\ref{fig:cluster_results}.

An outlier worth addressing is that of the Home graph.  Figure~\ref{fig:cluster_results} shows that in both the two dimensional reduced space, as well as the clustered space, this graph is closer to the stylized graphs than the real-world graphs.  We note that this graph is a collection of largely disconnected components and has low node degree with low degree variance and therefore appears similar to a regular graph.

\section{Conclusion}

We have presented a comparison of multiple synthetic social contact networks with real network data from the literature as well as multiple stylized graph generation methods, along several structural measures.

We conclude that the synthetic population-based method for generating graphs has quantifiable advantages over stylized methods, particularly in its ability to approximate the characteristics of real world graphs in a limited number of domains.  The reliability of this conclusion is qualified however by some room for improvement in experimental design and further research to either corroborate or erode this conclusion.  

To bolster the reliability of these results, future research should aim to include a larger and more varied assortment of real world graphs from both the social network domain as well as others for comparison on the basis of multiple types of underlying activity.  Moreover, future research should aim to be inclusive of other graph generation methods such as ``Structure-driven Models'' (e.g. Kronecker, dK-Graphs) and Intent-driven Models (e.g. Random Walk and Nearest Neighbor)~\cite{ornl_survey}.  In such research, attention should be given to scalability of graph feature computation.

The current results are also not completely evaluated: we have not attempted to define which features most powerfully cause separation or similarity of the graphs under analysis.  Furthermore, there are additional implementations of measures of entropy or similarity which may prove useful at capturing the nature of real world graphs versus synthetic ones.

Future research should strive to collect and evaluate data which represents \textit{multiple} underlying types of activity carried out by common vertices.  As described above, the G1 - G4 graphs are a combination of graphs in which a large majority of the nodes exist in all subgraphs.  Retrieving real world networks which remain faithful to such a complex set of data are difficult to acquire, but will be useful for comparing the utility of G1 - G4 against stylized counterparts for multi-channel graphs.

This work can also be extended to take into account dynamical measures, as discussed in the related work section, such as vulnerability and expansion, which provide insight into the realism of dynamical processes on networks.

\section*{Acknowledgments}
We acknowledge the Hume Center and the Biocomplexity Institute at Virginia Tech, as well as DARPA for their support in this research.


\bibliographystyle{IEEEtran}

\end{document}